\documentstyle[prl,aps,twocolumn,psfig]{revtex}

\begin{document}
\draft

\title{Auditory sensitivity provided by\\ self-tuned critical 
oscillations of hair cells}

\author{S\'ebastien Camalet$^*$, Thomas Duke$^{*\dagger\ddagger}$, Frank
J\"ulicher$^*$ and Jacques Prost$^*$}
\address{
$^*$Institut Curie, PhysicoChimie Curie, 
UMR CNRS/IC 168,\\ 26 rue d'Ulm,
75248 Paris Cedex 05, France\\
$^\dagger$Niels Bohr Institute, Blegdamsvej 17,
2100 Copenhagen, Denmark\\
$^\ddagger$Cavendish Laboratory, Madingley Road, Cambridge CB3 0HE, UK}
\maketitle
\pagenumbering{arabic}

\noindent
\begin{abstract}
{\bf We introduce the concept of {\em self-tuned criticality} as a 
general mechanism for signal detection in sensory systems. In the case of
hearing, we argue that active amplification of faint sounds is provided by
a dynamical system which is maintained at the threshold of an oscillatory
instability. This concept can account for the exquisite sensitivity of
the auditory system and its wide dynamic range, as well as its capacity
to respond selectively to different frequencies. A specific model of
sound detection by the hair cells of the inner ear is discussed. We show
that a collection of motor proteins within a hair bundle can generate 
oscillations at a frequency which depends on the elastic properties of 
the bundle. Simple variation of bundle geometry gives rise to hair cells
with characteristic frequencies which span the range of audibility.
Tension-gated transduction channels, which primarily
serve to detect the motion of a hair bundle, also tune each cell by
admitting ions which regulate the motor protein activity. By controlling
the bundle's propensity to oscillate, this feedback automatically 
maintains the system in the operating regime where it is most sensitive
to sinusoidal stimuli. The model explains how hair cells can detect
sounds which carry less energy than the background noise.}
\end{abstract}
\vspace{0.5cm}

Detecting the sounds of the outside world imposes stringent demands on
the design of the inner ear, where the transduction of acoustic
stimuli to electrical signals takes place \cite{huds89}. The hair
cells within the cochlea, which act as mechanosensors, must each be
responsive to a particular frequency component of the auditory input. 
Moreover, these sensors need the utmost sensitivity,
since the weakest audible sounds impart an energy, per cycle of
oscillation, which is no greater than that of thermal noise
\cite{devr49}. At the same time, they must operate over a wide range
of volumes, responding and adapting to intensities which vary by many
orders of magnitude. Clearly, some form of non-linear amplification is
necessary in sound detection. The familiar resonant gain of a passive
elastic system is far from sufficient for the required demands,
because of the heavy viscous damping at microscopic scales
\cite{gold48}. Instead, the cochlea has developed active amplificatory 
processes, whose precise nature remains to be discovered.

There is strong evidence that the cochlea contains force-generating
dynamical systems which are capable of executing oscillations of a
characteristic frequency 
\cite{huds97,dall92,craw85,howa88,bens96,zure81,prob90}.
In general, such a system exhibits a Hopf bifurcation \cite{stro94}:
as the value of a control parameter is varied, the behavior abruptly
changes from a quiescent state to self-sustained oscillations. When
the system is in the immediate vicinity of the bifurcation, it can act
as a nonlinear amplifier for sinusoidal stimuli close to the
characteristic frequency.  That such a phenomenon might occur in
hearing was first proposed by Gold
\cite{gold48}, more than 50 years ago. The idea was recently revived
by Choe, Magnasco and Hudspeth \cite{choe98}, in the context of a
specific model of the hair cell. No general analysis of the
amplification afforded by a Hopf bifurcation has been provided, 
however, and no theory has been advanced to explain how 
proximity to the bifurcation point might be ensured.

In this paper, we provide both a generic framework which describes the
known features of acoustic detection, and a detailed discussion of the
specific elements which could be involved in this detection. We first
derive the general resonance and amplification behavior of a dynamical
system operating close to a Hopf bifurcation and emphasize that such a
system is well-suited to the ear's needs.  In order for active
amplification to work reliably, tuning to the bifurcation point is
crucial. We introduce the concept of a {\em self-tuned Hopf
bifurcation} which permits the favorable amplificatory properties of a
dynamical instability to be obtained in a robust way. Self-tuning
maintains the system in the proximity of the critical point and is
achieved by an appropriate feedback mechanism which couples the output
signal to the control parameter that triggers the bifurcation. The
concept can explain several important features of the auditory sensor
such as the frequency selectivity, high sensitivity and the ability to
respond to a wide range of amplitudes.  It can also explain the 
intrinsic nonlinear nature of sound detection
\cite{jara93,cart99} and the occurrence of spontaneous sound emission
by the inner ear
\cite{zure81,prob90}. Furthermore, self-tuned criticality provides a
framework for understanding the role of noise in the detection
mechanism.  The amplificatory process, which involves a limited number
of active elements, introduces stochastic fluctuations, which adds to
those caused by Brownian motion. We show that the response to weak
stimuli can take advantage of this background activity.

The proposed existence of a self-tuned Hopf bifurcation raises
questions about the specific mechanisms involved: What is the physical
basis of the dynamical system?  How is the self-tuning realized?  It
might be expected that different organisms have evolved different
apparatus to implement the same general strategy. In this paper, we
restrict our specific discussion to the more primitive cochleae of
non-mammalian vertebrates. We propose a model of the hair cell of the
inner ear which accords with data from a wide variety of physiological
experiments.  The model incorporates a physical mechanism which allows
motor proteins to generate spontaneous oscillations \cite{juli97a}.
We find that molecular motors such as dyneins in the kinocilium or
myosins in the stereocilia are natural candidates for the force
generators involved in the amplification of hair-bundle
motion. Tension-gated transduction channels in the stereocilia serve
primarily to detect this motion, but also have a second function: by
admitting ions which regulate the motor protein activity, they provide
the self-tuning mechanism.

\section{Generic aspects}

{\bf Amplification and frequency filtering of a Hopf bifurcation.}  
We discuss the behavior of a dynamical system which is controlled
by a parameter $C$. Above a critical value, $C>C_c$, the
system is stable; for $C<C_c$ it oscillates spontaneously. At the
critical point (or Hopf bifurcation) $C=C_c$ the system shows
remarkable response and amplification properties which do not depend
on the physical mechanism at the basis of the bifurcation.  These
generic properties can be described as follows. Since we are
interested in the response to a periodic stimulus with frequency
$\nu=\omega/2\pi$, we express the hair-bundle deflection $x(t)$ by a
Fourier series 
$x(t)=\sum x_n e^{i n \omega t}$
with complex amplitudes $x_n=x_{-n}^*$. In the vicinity of the
bifurcation, the mode $n=\pm 1$ is dominant and the
response to an externally-applied sinusoidal stimulus force $f(t)=f_1
e^{i\omega t} + f_{-1} e^{-i\omega t}$ can be expressed in terms
of a systematic expansion in $x_1$. Symmetry arguments
(see Appendix A) imply that the first nonlinear term is
cubic:
\begin{equation}
f_1 = {\cal A}\; x_1 + {\cal B}\;  |x_1|^2 x_{1} + ...
 \label{eq:hopf}
\end{equation}
where ${\cal A}(\omega,C)$ and ${\cal B}(\omega,C)$ are two complex
functions. The bifurcation point is characterized by the fact that
${\cal A}$ vanishes for the critical frequency, ${\cal
A}(\omega_c,C_c)=0$. For $C<C_c$ and no external force, the system
oscillates with $|x_1|^2 \simeq \Delta^2 \; (C_c-C)/C_c$, where $\Delta$
is a characteristic amplitude.  For $C=C_c$ the response to a stimulus
at the critical frequency has amplitude
\begin{equation}
|x_1| \simeq |{\cal B}|^{-1/3} |f_1|^{1/3} \quad .
\end{equation}
This represents an amplified response
at the critical frequency with a gain
\begin{equation}
r=\frac{|x_1|}{|f_1|} \sim |f_1|^{-2/3} \label{eq:gain}
\end{equation}
that becomes arbitrarily large for small forces.

If the stimulus frequency differs from the critical frequency, 
the linear term in Eq. (\ref{eq:hopf}) is non-zero and can be
expressed to first order as
${\cal A}(\omega,C_c)\simeq A_1\; (\omega-\omega_c)$. The dramatic
amplification of weak signals, implied by Eq. (\ref{eq:gain}), is
maintained as long as this term does not exceed the cubic term
in Eq. (\ref{eq:hopf}). If the frequency mismatch increases such
that $|\omega-\omega_c|\gg |f_1|^{2/3}|{\cal B}|^{1/3}/|A_1|$,
the response becomes linear 
\begin{equation}
|x_1| \simeq \frac{|f_1|}{|(\omega-\omega_c)A_1|}
\end{equation}
i.e. the gain is independent of the strength of the stimulus.

\begin{figure}
\centerline{\psfig{figure=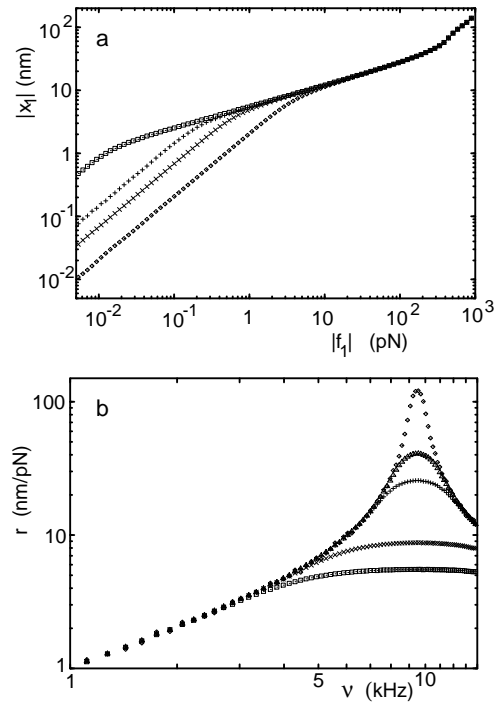,width=6.5cm}}
\bigskip
\caption{ Response to external forces near a Hopf bifurcation
(a) Amplitude $|x_1|$ as a function of force $|f_1|$ at various driving
frequencies $\nu$ ($\Diamond$ 2 kHz, $\times$ 5 kHz, $\Box$ 10 kHz,
$+$ 13 kHz). (b) Gain $r$ as a function of frequency $\nu$ for different
amplitudes $|f_1|$ ($\Diamond$ 0.01 pN, $\triangle$ 0.05 pN, $+$ 0.1 pN,
$\times$ 0.5 pN, $\Box$ 1 pN).
While the form of these curves is generic, the numerical values of 
force, amplitude and frequency depend on the physical nature of the
dynamical system. The values given here correspond to the specific model
model of a hair cell discussed in the text, with parameters chosen to 
give a critical frequency of approximately 10\,kHz.}
\label{fig:response}
\end{figure}

Thus the Hopf resonance acts as a sharply tuned high-gain amplifier
for weak stimuli, and as a low-gain filter for strong stimuli. This
generic behavior is illustrated in Fig. 1 with data obtained by
numerical simulation. Laser interferometry measurements of the motion
of the basilar membrane, when the live cochlea is stimulated by pure
tones, display strikingly similar features \cite{rugg92,rugg97}.  In
particular, the peak response as a function of force amplitude has
been demonstrated to obey a power law $|x_1|\sim |f_1|^{0.4\pm 0.2}$
\cite{rugg97}. This strongly suggests that the membrane is being
driven by a dynamical system which is poised at a Hopf bifurcation.

{\bf Self-tuned critical oscillations.}  How does the system come to
be so precisely balanced at the critical point?\footnote{Self-tuning
to a coexistence point has been discussed 
previously in certain dynamic first-order transitions \cite{coul89}.}
The control parameter
must be tuned to $C\simeq C_c$, otherwise the nonlinear amplification
is lost. Moreover, the value of $C_c$ differs for hair cells with
different characteristic frequency. We propose a feedback mechanism
which allows the dynamical system to operate {\em automatically} close
to the bifurcation point, whatever its characteristic frequency.
Without loss of generality, we assume that the control parameter
decreases as long as the system does not oscillate.  After some time,
critical conditions are reached and spontaneous oscillations
ensue. The onset of oscillations triggers an increase of the control
parameter which tends to restore stability. Hence the system
converges to an operating point close to the bifurcation point. To
illustrate this general idea, we consider the following simple
feedback which changes $C$ in response to deflections $x$:
\begin{equation}
\frac{1}{C}\frac{\partial C}{\partial t} = \frac{1}{\tau}\left(
\frac{ x^2}{\delta^2} -1\right) \label{eq:st}
\end{equation}
where $\delta$ is a typical amplitude.  If no external force is
applied, this feedback, after a relaxation time $\tau$, tunes the
control parameter to a value $C_{\delta}$ for which spontaneous
oscillations with $|x_1| \simeq \delta$ occur.  If $\delta$ is small
compared to the characteristic amplitude $\Delta$, this is on the
oscillating side close to the bifurcation, $|C_{\delta}-C_c|/C_c\simeq
(\delta/\Delta)^2$. Two modes of signal detection are possible: (i)
For transient stimuli short compared to $\tau$ the system operates at
$C_{\delta}$. The amplitude $|x_1|$ shows the characteristic nonlinear
response discussed above.  (ii) For stimuli sustained over longer
times, self-tuning maintains $|x_1|
\simeq \delta$ constant for different stimulus amplitudes.  This
effect of the feedback represents a perfect adaptation
mechanism. However, in the presence of noise, phase-locking of the
response (to be discussed later) occurs as soon as an external
stimulus is applied, and this can be detected.

\section{Model}

{\bf Mechanosensitivity and self-tuning mechanism provided by
transduction channels.}  We demonstrate the general principles
introduced above by devising a specific model for the amplification of
acoustic stimuli by hair bundles in non-mammalian vertebrates.  
A schematic representation of a hair bundle is shown in Fig. 2.
It consists of several stereocilia and a single kinocilium 
\cite{tiln83,jaco90}. Transduction of hair bundle deflection to a
chemical signal occurs via channels located near the tip of each
stereocilium.  Tip links which connect neighboring stereocilia are
believed to be the gating springs of the transduction
channels. If the hair bundle is deflected, tension in the tip links
triggers the opening of the channels. The subsequent influx of ions
(principally $\rm K^+$, but also $\rm Ca^{2+}$) causes a corresponding
change of the membrane potential of the hair cell which, in turn,
generates a nervous signal.

The mechanosensor can be used for self-tuning, as well as signal detection.
Many physiological processes are regulated by ionic concentrations, so it is
natural to identify the $\rm Ca^{2+}$ concentration with the control
parameter $C$.  We assume that $C$ decreases if the transduction
channels are closed, owing to the action of pumps in the cell membrane.
When the hair bundle is deflected by $x$, the transduction channels open 
with probability $P_o(x)$.  We therefore characterize the mechanosensor 
by the equation
\begin{equation}
\frac{\partial C}{\partial t} = - \frac{C}{\tau} + J_o P_o(x) 
\label{eq:dyn_C}
\end{equation}
where $J_o$ is the $\rm Ca^{2+}$ flux through open transduction
channels.  Note that this equation provides self-tuning. In
this case it replaces the more simple but less realistic
Eq. (\ref{eq:st}).  For our numerical examples, we use a two-state
model for the channels with
\begin{equation}
P_o(x)  = \frac{1}{1+A e^{-x/\delta}} \label{eq:Po}
\end{equation}
where $(1+A)^{-1} \ll 1$ is the probability that a channel is open when the
hair cell is quiescent and $\delta$ is the
characteristic amplitude of motion to which the system is sensitive.
For a sufficiently long relaxation time $\tau$, the
slow variation $C_0$ of $C\simeq C_0(t)+C_1 e^{i\omega t}$ 
can be separated from the small-amplitude oscillations, giving
\begin{equation}
\partial_t C_0  \simeq  -\frac{C_0}{\tau}+J_o \tilde P_o(|x_1|^2)
\label{eq:stOm}
\end{equation}
where $\tilde P_o=\int_0^{1/\nu} dt\; P_o(x_1 e^{i\omega
t}+x_{-1}e^{-i\omega t})$ is the averaged probability of channel
opening in the presence of oscillations, which increases monotonically
with their amplitude $x_1$. For physically relevant parameter 
values, the system reaches a steady state close to
the bifurcation point, independent of the initial value of $C$.

{\bf Oscillations generated by molecular motors.}  We still have to
specify the nature and location of the oscillator within the hair
bundle. It has been suggested that the transduction channels might be
the source of the instability
\cite{howa88,choe98}; or, that myosin motors 
in the stereocilia might generate the force necessary to move the
bundle \cite{maca80,manl97}. We propose a third possibility: the
kinocilium could vibrate using its internal dynein motors.
Recently, a simple physical mechanism has been proposed which allows
motor proteins operating in collections to generate spontaneous
oscillations by traversing a Hopf bifurcation
\cite{juli97a,juli97b}. Typically, motors move along cytoskeletal filaments
and elastic elements oppose this motion. In this case, two
possibilities exist: The system either reaches a stable balance
between opposing forces, or it oscillates around the balanced state.
Three time scales characterize this behavior: The relaxation time
$\lambda/K$ of passive relaxation, where $\lambda$ is the total
friction and $K$ is the elastic modulus; and the times $\omega_1^{-1}$
and $\omega_2^{-1}$, where $\omega_1$ is the kinetic rate at which a
motor detaches from a filament, and $\omega_2$ is the attachment rate.
An explicit solution of a simple model is derived in Appendix B.
 We find that for an appropriate value of a control parameter
$C$, which is related to the ratio $\omega_1/\omega_2$, a Hopf
bifurcation occurs with critical frequency
\begin{equation}
\omega_c \simeq
\left(\frac{K\alpha}{\lambda}\right)^{1/2} \quad ,
\end{equation} 
which is the geometric mean of the passive
relaxation rate $K/\lambda$ and the typical ATP hydrolysis rate
$\alpha = \omega_1+\omega_2$.  The above identification of $C$ with
the $\rm Ca^{2+}$ concentration is consistent with the fact that $\rm
Ca^{2+}$ regulates motor protein activity \cite{walc94}.  The system
oscillates if the elastic modulus does not exceed a maximal value
$K_{\rm max}\simeq k_0 N$, where $k_0$ is the crossbridge elasticity
of a motor and $N$ is the total number of motors.  The maximal
frequency, obtained when $K=K_{\rm max}$, can be significantly higher
than the ATP hydrolysis rate $\alpha$.

\begin{figure}
\centerline{\psfig{figure=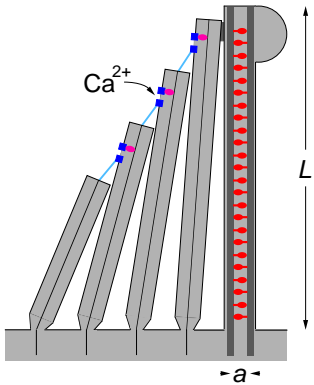,width=4.5cm}}
\bigskip
\caption{A hair bundle consists of a single kinocilium and tens to hundreds
of stereocilia. The kinocilium contains dynein motors (red). The
stereocilia contain transduction channels (blue) which are gated by the
tension in the tip links (cyan); the steady-state tension is maintained
constant by adaptation motors (magenta). In our model, the
kinocilium is the active part of the mechanoreceptor and the stereocilia
act as the detection apparatus. Feedback control of the active 
amplificatory process involves the influx of $\rm Ca^{2+}$ ions through
the transduction channels; the ions diffuse into the kinocilium and
regulate the generation of force by the motor proteins.}
\label{fig:hair_bundle}
\end{figure}

{\bf Characteristic frequency of a vibrating kinocilium.}  The
kinocilium is a true cilium containing a cylindrical arrangement of
microtubule doublets and dynein motors. Since cilia have the
well-established tendency to beat and vibrate with frequencies from
tens of Hz up to at least 1$\,$kHz \cite{gibb81,kami89}, the
kinocilium is a natural candidate to be responsible for the Hopf
bifurcation. A simple two-dimensional model can be used to discuss the
main physical properties of a vibrating cilium near a Hopf bifurcation
\cite{cama99}. In this model, motors induce the bending of a pair of
elastic filaments separated by a distance $a$ (corresponding to the
distance between neighboring microtubule doublets in the axoneme).  An
isolated kinocilium of length $L$ and bending rigidity $\kappa$, fixed
at the basal end but free at its tip, will vibrate in a wave-like mode
with wavelength $\Lambda \simeq 4 L$.  A typical displacement $z$ of
the motors along the filaments leads to bending of the filament pair
and a deflection of the tip by a distance $x \simeq zL/a$. The elastic
bending energy is of order $(\kappa/L^3)x^2\simeq (\kappa/a^2 L)
z^2$. Therefore, the total elastic modulus experienced by the motors
is given by
\begin{equation}
K \simeq \frac{\kappa}{L a^2} \quad. \label{eq:K}
\end{equation}
The viscous energy dissipation per unit time due to motion of the kinocilium
is of order $\eta L (\partial_t x)^2\simeq (\eta L^3/a^2)(\partial_t
z)^2$. Therefore, the total friction experienced by the
motors can be written as
\begin{equation}
\lambda \simeq \frac{\eta L^3}{a^2} + \lambda_0 \rho L \quad
,\label{eq:lambda} 
\end{equation}
where $\lambda_0$ describes 
the dissipation within the kinocilium per motor and $\rho=N/L$ denotes
the number of dynein motors per unit length along the axoneme. If
internal friction can be neglected, i.e. if $L \gg L_0 = (\lambda_0
a^2 \rho/\eta)^{1/2}$, the frequency of a vibrating cilium at the
bifurcation point is given by
\begin{equation}
\omega_c \simeq \left( \frac{\kappa \alpha}{\eta}\right)^{1/2} \frac{1}{L^2}
\quad. \label{eq:om_c}
\end{equation} 
Using typical values $\lambda_0\simeq 10^{-9}
\,{\rm kg/s}$, $a\simeq 20\,{\rm nm}$, $\rho \simeq 5
\; 10^{8}\,{\rm m^{-1}}$ and $\eta \simeq 10^{-3}\,{\rm kg/ms}$, 
we find $L_0\simeq 200\,{\rm nm}$, shorter than typical kinocilia.
Using $\alpha \simeq 10^3\,{\rm s^{-1}}$, $k_0 \simeq 10^{-3}\,{\rm N/m}$
and $\kappa \simeq 4 \; 10^{-22}\,{\rm Nm^2}$ (the bending
rigidity of 20 microtubules), 
the frequency range between 100$\,{\rm Hz}$ and 10$\,{\rm kHz}$ 
can naturally be spanned by changing the length of the kinocilium
between $1\,\mu{\rm m}$ and $10\,\mu{\rm m}$.

The above argument neglects the contribution of the stereocilia to
elasticity and motion. The elastic response of stereocilia to hair
bundle deflections has been measured \cite{craw85,howa86}. It can be
well described by an angular spring at the base of each stereocilium
which contributes an elastic energy per stereocilium of the order of
$k_s (x/L)^2$, where $k_s$ is an angular elastic modulus. The kinocilium
length $L$ in a hair bundle is approximately inversely proportional to
the number $N_s$ of stereocilia \cite{tiln83}, and we write $N_s\simeq
l_s/L$, where $l_s\simeq 10^{-4}\,{\rm m}$ is the total length of
stereocilia.  With this assumption, we find an additional elastic
modulus
\begin{equation}
K_s\simeq \frac{l_s k_s}{La^2}
\end{equation}
contributed by the stereocilia to $K$.  The contribution of
stereocilia to the friction coefficient $\lambda$ can be estimated as
\begin{equation}
\lambda_s \simeq \eta l_s L^2/a^2 \quad .
\end{equation}
The measured value of $k_s$ \cite{howa86}
indicates that $K_s$ dominates the contribution to $K$ given by
Eq. (\ref{eq:K}). Similarly, since $l_s>L$, $\lambda_s$ should dominate
friction. In this case, we expect
\begin{equation}
\omega_c\simeq\left(\frac{k_s \alpha }{\eta L^3}\right )^{1/2} 
\end{equation}
and the range of frequencies is somewhat reduced.

If the stiffness of the ensemble of stereocilia greatly exceeds that
of the kinocilium, a new situation arises. Since the kinocilium is
attached to the stereocilia at its tip \cite{jaco90}, movement of the
tip is strongly reduced and the kinocilium preferentially vibrates in
a mode with wavelength $\Lambda\simeq 2 L$ for which the relation given by
Eq. (\ref{eq:om_c}) is again valid.

{\bf Adaptation motors.}  Finally, we note that in order to obtain a
robust self-tuning, the feedback mechanism must be sensitive only to
the oscillation amplitude $x_1$ and not to a stationary displacement
$x_0$; we assumed in Eq. (\ref{eq:stOm}) that $\tilde P_o$ is
independent of $x_0$. The transduction mechanism of stereocilia and
their tension-gated channels does indeed have this property: It is
well known that an ATP-dependent adaptation mechanism \cite{huds94}
exists which removes the dependence of the channel current on a
constant displacement $x_0$. It is widely believed that this
adaptation involves the motion of myosin motors, which maintains
constant the steady-state tension in the tip links that control the
transduction channels \cite{gill97}. Therefore, the stereociliar
transduction mechanism has precisely the required properties to be
used as a feedback signal for self-tuning the bifurcation.

\section{Simulation}
The above analysis does not include the effects of noise.  Brownian
motion is one source of fluctuations in the movement of the hair
bundle \cite{denk89}. Another source of noise is caused by stochastic
fluctuations in motor protein force as dynein molecules bind to, and
detach from the microtubules in the kinocilium. In order to
investigate the consequences of this randomness, we performed a Monte
Carlo simulation of the two-state model described in \cite{juli97a},
using a realistic number of motor proteins. The motors, when attached,
experienced a potential of amplitude $U$ and period $l$ (the
corresponding crossbridge elasticity is $k_0 = U/l^2$).  The
attachment rate was constant and independent of position,
$\omega_2=\alpha$. Detachment was localized to a region of width
$0.1\,l$, centred on the potential minimum and the detachment rate
$\omega_1$ was regulated by the $\rm Ca^{2+}$ concentration $C$. We chose
$\omega_1/\omega_2 = (C_q/C)^3$, where $C_q$ is the steady-state $\rm
Ca^{2+}$ concentration in a quiescent hair cell; the precise
functional dependence is unimportant as long as $\omega_1/\omega_2$
decreases monotonically with increasing $C$ in a fairly sensitive way.
We simulated systems with $N$ = 1000-4000 motors, representing hair
cells with different kinocilium length and stereociliary number.

\begin{figure}
\centerline{\psfig{figure=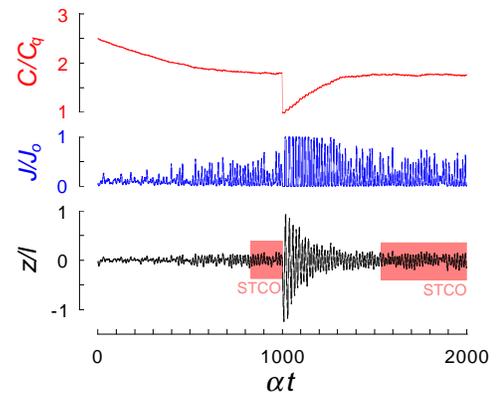,width=6.5cm}}
\bigskip
\caption{Self-tuning of a hair bundle. If the ${\rm Ca^{2+}}$
concentration $C$ within the cell is artificially high, the hair
bundle is initially quiescent; but as ${\rm Ca^{2+}}$ ions are pumped
from the cell, it gradually begins to oscillate with small
amplitude. If $C$ is suddenly artificially lowered, the hair bundle
becomes unstable and executes high-amplitude spontaneous oscillations.
Since these movements open the transduction channels, the influx $J$
of ${\rm Ca^{2+}}$ ions increases; the consequent change in $C$
regulates the motors and diminishes the amplitude of the oscillation
until it almost disappears. In the steady state, the bundle executes
{\em self-tuned critical oscillations} (STCO).  For this simulation,
we assumed that the probability of transduction channels opening was
$P_o(z) = (1+10e^{-20z/l})^{-1}$, where $z$ is the motor displacement,
and that the time constant for equilibration of the ${\rm Ca^{2+}}$
concentration was $\tau = 1000/\alpha$.}
\label{fig:STCO}
\end{figure}

{\bf Self-tuning and the characteristic frequency of spontaneous
oscillations.}  The self-tuning of a hair cell to the vicinity of the
bifurcation, where small-amplitude spontaneous oscillations occur, is
demonstrated in Fig. 3.  When a change of the internal $\rm Ca^{2+}$
concentration is imposed, the system is transiently perturbed; but
after an interval of time of order $\tau$, it returns to the same
steady state. Hair cells with different numbers of motors acquire
different internal concentrations of $\rm Ca^{2+}$, in order to adjust
the motor detachment rate in such a way that the system approaches the
critical point. The spontaneous oscillations of three different hair
bundles are shown in Fig. 4. Note that the characteristic frequency is
approximately proportional to the inverse-square of the number of
motors, as expected from Eq. (\ref{eq:om_c}), and that it can exceed
the typical ATP cycle rate $\alpha$ when the total number of motors is
small (short kinocilium).  All three hair cells execute spontaneous
oscillations with a similar amplitude, as expected from our arguments
above. However, the noise introduces a significant new effect: The
oscillations are irregular. The incoherence of the phase of the
oscillation is evident in the Fourier transform of the displacement,
which exhibits a broad peak centred on the characteristic frequency.

\begin{figure}
\centerline{\psfig{figure=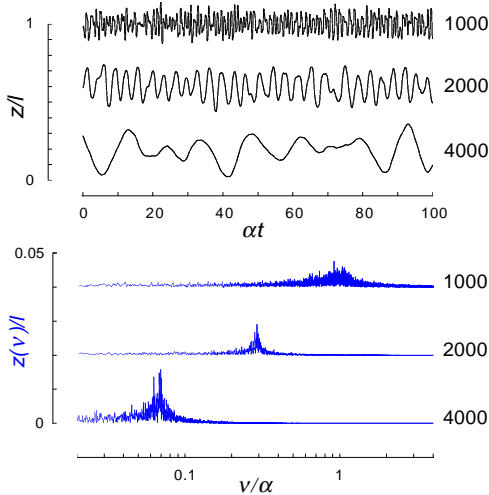,width=6.5cm}}
\bigskip
\caption{Self-tuned critical oscillations of systems comprising
$N = 1000, 2000$ and 4000 motors. In each case, the oscillation has
small amplitude $z \simeq 0.1\,l$ and is irregular, as can be seen
from the broadness of the Fourier spectrum.  For these simulations we
chose parameters $k_0/\lambda \alpha = 4 \;10^7 N^{-3}$ and $K/\lambda
\alpha = 4 \;10^{13} N^{-4}$, which correspond with the scaling
dependence of the spring constant and friction coefficient on
kinocilium length in Eqs. (\ref{eq:K}-\ref{eq:lambda}), and also with
the order of magnitude estimates of the physical parameters, given in
the text.}
\label{fig:stco}
\end{figure}

{\bf Dynamic response to a tone at the characteristic frequency.}  The
response of a self-tuned hair bundle to a sinusoidal force with a
frequency approximately equal to the bundle's characteristic frequency
is illustrated in Fig. 5. For weak stimuli,the amplitude of the oscillation
does {\em not} increase with the amplitude $|f_1|$ of the applied force;
this is because the small response to the stimulus is masked by the
noisy, spontaneous motion. Instead, the {\em phase} of the hair-bundle
oscillation becomes more regular; as it does so, a peak emerges from the
Fourier spectrum at the driving frequency. The height of the peak grows
approximately as  $|f_1|^{1/3}$ for intermediate values of $|f_1|$, and
approximately linearly for very weak stimuli, and also for very strong
stimuli 
(for which the response is essentially passive). Thus the Fourier component 
of the hair-bundle displacement at the driving frequency responds to the 
stimulus in the generic manner discussed above.

\begin{figure}
\centerline{\psfig{figure=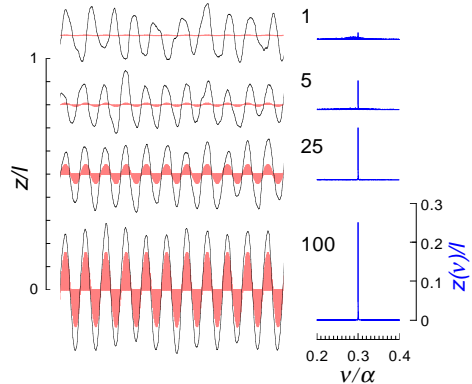,width=6.5cm}}
\bigskip
\caption{Response of a system with $N = 2000$ motors to a sinusoidal
force at frequency $\nu = 0.3 \alpha$, close to the hair
bundle's characteristic frequency. Curves are marked by
the dimensionless amplitude of the force, $|f_1|/f_{mot}$,
where $f_{mot}=U/l$ is the force produced by a motor molecule.
Note that a force equal to that of a single motor is sufficient
to elicit a response in the Fourier spectrum. Curves shaded red
are the responses of an equivalent, passive hair bundle (i.e. a
bundle with identical mechanical properties but no force 
generators).}
\label{fig:sinf}
\end{figure}

{\bf Phase-locking of the nervous signal.}  The flow of ions through
the transduction channels depolarizes the cell membrane which, in
turn, opens voltage-gated channels at the base of the hair cell and
generates a synaptic current \cite{huds89}. This sequence of events
happens fast enough for variations at the synapse faithfully to
reflect the hair-bundle motion at frequencies below
1\,kHz. Information about the auditory stimulus is subsequently passed
along the auditory nerve in the form of a spike train.  Simplifying
this transduction process, we assume that a spike is elicited whenever
the transduction channel current $J$ passes a threshold value of $0.5
J_o$. The resulting nervous response is shown in Fig. 6.  In the
absence of a stimulus, the self-tuned critical oscillations of the
bundle cause the nerve to fire stochastically, at a low rate. When a
weak sinusoidal stimulus is applied at the characteristic frequency,
the firing rate does not increase above the spontaneous rate, but the
spike train becomes detectably phase-locked to the stimulus.  The
degree of phase-locking increases rapidly as the amplitude of the
stimulus increases, reflecting the growing regularity of the
hair-bundle motion.  It is only when the neural response is almost
completely phase-locked that the firing rate begins to
rise. Eventually, for strong stimuli, the spike rate saturates at the
stimulus frequency. This behavior is strikingly similar to that which
is observed experimentally. In particular, it is well known that the
threshold for phase-locking in the auditory nerve fiber is 10-20 dB
lower than the threshold at which the firing rate begins to rise
\cite{hill84,koep97}.

\begin{figure}
\centerline{\psfig{figure=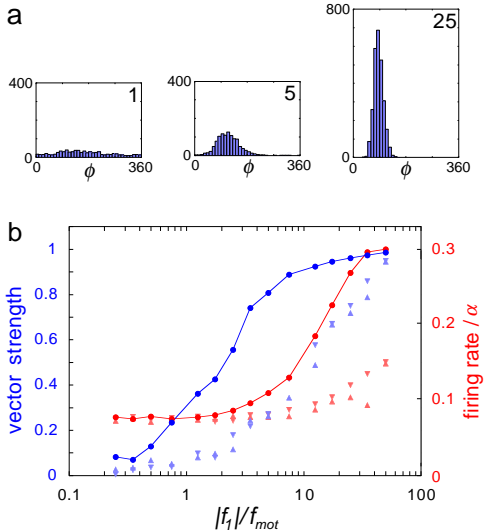,width=6.5cm}}
\bigskip
\caption{(a) Histogram of the phase $\phi$ of the driving force at the instants
when nervous spikes are generated, for various values of the force
amplitude (marked by the value of $|f_1|/f_{mot}$).  (b) Degree of
phase-locking (blue; expressed as the vector strength $\int d\phi\;
p(\phi)e^{i\phi}$) and firing rate (red), as a function of the
driving force. The frequency selectivity of the bundle can be
appreciated by comparing the neural response to a stimulus with
frequency close to the characteristic frequency, $\nu = 0.3 \alpha \,
(\bullet)$, with the response to other frequencies, $\nu = 0.15 \alpha
\,(\bigtriangledown) $ and $\nu = 0.6 \alpha \, (\triangle)$.}
\label{fig:phase_locking}
\end{figure}

\section{Discussion}

The self-tuned critical oscillator which we have introduced as a
system for signal detection has characteristic amplification properties 
which are generic and do not depend on the choice of a specific model.
The oscillation frequency, however, depends on the physical mechanism
involved. In our model of non-mammalian hair cells, the frequency is
determined by the geometry of the hair bundle. A simple
morphological gradient along the basilar membrane would endow the
ear with the ability to analyze a wide range of frequencies.
Experimentally, it is well established that the height of
hair bundles progressively increases along the cochlea, and that
concurrently the characteristic frequency of the hair cells declines
\cite{tiln83,holt83}. Our proposition that the kinocilium is likely to 
play an active role in non-mammalian vertebrate hair cells suggests
experiments which study the motility of the kinocilium and its
potential for generating oscillations. If the kinocilium is the only
source of oscillations, its removal should suppress hair-bundle
vibrations. If the self-tuning mechanism is removed instead (by
cutting the tip links, for example) our model predicts that the
kinocilium should exhibit stronger spontaneous oscillations. Careful
control of the extracellular ion concentrations, such as $\rm Ca^{2+}$
would be essential in such experiments.

The simulations of our model reveal how a hair bundle can achieve its
remarkable sensitivity to weak stimuli.  By profiting from the
periodicity of a sinusoidal input, and measuring phase-locking rather
than the amplitude of response, the mechanosensor can detect forces
considerably weaker than those exerted by a single molecular motor (if
the bundle were a simple, passive structure, its response to such
forces would be smaller than its Brownian motion).  An important
implication of this detection mechanism is that even though the hair
cell selects a certain frequency, the signal must still be encoded by
the interval between spikes elicited in the auditory
nerve. Paradoxically, the stochastic noise caused by the motor
proteins serves a useful purpose. It ensures that the self-tuned
critical oscillations of the hair bundle are incoherent, so that the
pattern of spontaneous firing in the nerve is irregular. Against this
background, the regular response to a periodic stimulus can easily be
detected.\footnote{The benefits of noise have been discussed
in a variety of situations such as those involving stochastic
resonance \cite{wies95}.}

Another beneficial feature of noise arises from the fact that weak 
stimuli do not increase the amplitude of oscillation above the
spontaneous amplitude. Thus the ${\rm Ca^{2+}}$ concentration remains
constant, the hair bundle stays in the critical regime, and active
amplification can be sustained indefinitely.  Stronger stimuli cause
the system to drift away from the critical point, so that the degree
of amplification diminishes over time. It is well known that both the
perception of loudness \cite{scha83} and the firing rate of the
auditory nerve \cite{youn73} decrease over a period of a few seconds,
when a stimulus of moderate intensity is maintained. This phenomenon,
which is usually referred to as `adaptation', is consistent with our
self-tuning mechanism.  Following the sustained presentation of a loud
stimulus, the spontaneous firing rate diminishes and the threshold to
weak stimuli is augmented \cite{youn73}.  Such `fatigue' is also
accounted for by our self-tuning mechanism. 

Since self-tuning positions the system slightly on the oscillating
side of the critical point, self-tuned criticality provides a natural
explanation for otoacoustic emissions
\cite{zure81,prob90}. In its normal working state, the inner ear would
generate faint sounds with a broad range of frequencies. If the
feedback mechanism were to fail in certain cells, the spontaneous
oscillations could become large enough for distinct tones to be
emitted.

A self-tuned Hopf bifurcation is ideal for sound detection because it
provides sharp frequency selectivity and a nonlinear gain which
compresses a wide range of stimulus intensities into a narrow range of
response.  We therefore believe that the concept applies to all
vertebrate hearing systems, and potentially to other mechanoreceptor
systems.  Kinocilia are typically absent in mammalian cochlea, and we
suggest that their force-generating role has been assumed by the outer
hair cells.  Self-tuning of these motile cells might be realized by
a mechanism similar to that presented here, using transduction channels 
in their hair bundles. It could, however, work very differently; 
for example, it might involve feedback from the inner hair cells,
via the efferent nervous system.

Tuning to the proximity of a critical point is likely to be a general
strategy adopted by sensory systems. Simple molecular receptors
\cite{duke99}, as well as the physiological sensors of higher
organisms, can enhance their response to weak stimuli in this way. We
propose that the physics of self-tuned criticality is the `central
science of transducer physiology' spoken of by Delbr\"uck
\cite{delb70}.

\vspace{1.0cm}
{\bf Acknowledgements:} We thank P. Martin and C. Petit for useful
discussions, and A.F. Huxley for referring us to the work of
T. Gold. T. Duke is grateful for the hospitality of Institut Curie and
the Niels Bohr Institute, and acknowledges the support of the Royal
Society. After submission of our manuscript
we learned that Eguiluz, Ospeck, Choe, Hudspeth and Magnasco have 
independently described the generic response of a system near a Hopf 
bifurcation; we thank them for communicating their unpublished results.

\appendix

\section{Generic behavior at a Hopf-bifurcation}

\subsection{Nonlinear relation between periodic stimulus and
displacements}

We are interested in the response $x(t)$ of a nonlinear system to a
periodic stimulus force $f(t)$. If only one frequency
$\nu=\omega/2\pi$ is present we use the Fourier expansions
\begin{eqnarray}
f(t)& = & \sum_{n=-\infty}^{\infty} f_n e^{i n \omega t} \\
x(t)& = & \sum_{n=-\infty}^{\infty} x_n e^{i n \omega t} \quad ,
\end{eqnarray}
where the complex coefficients $x_n$ and $f_n$ obey $x_{n}=x_{-n}^*$
and $f_n = f_{-n}^*$. This representation implies that we focus on the
limit cycle solution and ignore all transient relaxation phenomena.
We consider the class of systems for which the force at a given time
depends in a nonlinear way on the history of the displacements $x(t)$
alone; as we will discuss in section D more complex cases do not
change the basic properties. In this situation, the relation between
$x$ and $f$ can be expressed as a systematic expansion of the force
amplitudes $f_n$ in the amplitudes $x_n$:
\begin{eqnarray}
f_k = F^{(1)}_{kl} x_l &+& F^{(2)}_{klm} x_l x_m \nonumber \\ 
&+& F^{(3)}_{klmn} x_l x_m x_n
+ O(x^4) \quad , \label {eq:expnl}
\end{eqnarray}
where the expansion coefficients $F^{(n)}_{k,k_1,..,k_n}$ are
symmetric with respect to permutations of the indices $k_1 .. k_n$.
The limit cycle solutions are invariant with respect to translations
in time $t\rightarrow t+\Delta t$. Under these transformations the
amplitudes change as $x_n \rightarrow x_n e^{i n\omega \Delta t}$ and
$f_n \rightarrow f_n e^{i n \omega \Delta t}$. Inspection of
Eq. (\ref{eq:expnl}) shows that the time translation symmetry allows
only for those terms $F^{(n)}_{k,k_1,..,k_n} x_{k_1}..x_{k_n}$ for
which $k = k_1+...+ k_n$. For all other cases $F^{(n)}_{k,k_1,..,k_n}$
must vanish which significantly restricts the number of terms.

\subsection{Hopf bifurcation}

The nonlinear system exhibits spontaneous oscillations and a
Hopf-bifurcation if nontrivial solutions to Eq. (\ref{eq:expnl}) with
$x_n\neq 0$ exist in the case where all $f_k=0$, i.e.  if no stimulus
force is applied. Without loss of generality, we consider here an
instability of the mode $x_1$. In this case, the dominant terms
allowed by symmetry read ($f_k=0$)
\begin{eqnarray}
0 & \simeq & F^{(1)}_{11} x_1 + 2 F^{(2)}_{1,2,-1} x_{-1} x_2 \nonumber \\
 &+& 6 F^{(3)}_{1,1,1,-1} x_1^2 x_{-1} + 6 F^{(3)}_{1,1,2,-2} x_2 x_{-2} x_1
\label{eq:f1}\\ 
0 & \simeq & F^{(2)}_{22} x_2 + 2 F^{(2)}_{211} x_1^2 \quad .
\label{eq:f2}
\end{eqnarray}
Eq. (\ref{eq:f2}) determines $x_2
\simeq -2 (F^{(2)}_{211}/{F_{22}^{(2)}}) x_1^2$.  Inserting this
relation in Eq. (\ref{eq:f1}), we obtain to lowest order
\begin{equation}
0 \simeq {\cal A} x_1 + {\cal B} |x_1|^2 x_{1} \quad , \label{eq:nonlin}
\end{equation}
where ${\cal A} \equiv F^{(1)}_{11}$ and 
${\cal B}\equiv 3 F^{(3)}_{1,1,1,-1} - 4 F^{(2)}_{211}
F^{(2)}_{1,2,-1}/F_{22}^{(2)}$.

The coefficients ${\cal A}(\omega,C)$ and ${\cal B}(\omega,C)$ are
complex and in general depend on frequency $\omega$ and a control
parameter which we denote by $C$. A Hopf bifurcation occurs at a
critical point $C=C_c$ at which ${\cal A}$ vanishes for a frequency
$\omega_c$, i.e. ${\cal A}(\omega_c,C_c)=0$. This can be demonstrated
as follows: A spontaneously oscillating solution satisfies
\begin{equation}
|x_1|^2 = -\frac{{\cal A}}{\cal B}  \label{eq:amp}
\end{equation}
Note, that such a solution can only exist if ${\cal A}/{\cal B}$ is
real and negative. At the bifurcation point, ${\cal A}=0$ and ${\cal
A}/{\cal B}$ is therefore real for $\omega=\omega_c$, however the
corresponding amplitude $|x_1|^2$ vanishes. In the vicinity of this
point we expect to find solutions with finite amplitude. We use the
expansion
\begin{equation}
{\cal A}(\omega,C)\simeq  (\omega-\omega_c) A_1 +  (C-C_c) A_2
\end{equation}
where $A_1$ and $A_2$ are complex coefficients and we neglect higher
order terms.  Spontaneous oscillating solutions exist only if ${\cal
A}/{\cal B}$ is real. This condition is satisfied for a particular
frequency $\omega=\omega_s$ with
\begin{equation}
\omega_s=\omega_c+\frac{Im(A_2/{\cal B})}{Im(A_1/{\cal B})}(C_c-C) \quad .
\end{equation} 
The ratio $-{\cal A}/{\cal B}$ at this frequency $\omega_s$ changes
sign for $C=C_c$; here we assume without loss of generality that it is
positive for $C<C_c$. In this case, the system oscillates
spontaneously with an amplitude which according to Eq. (\ref{eq:amp})
behaves as $|x_1|^2=\Delta^2 (C_c-C)/C_c$, where
\begin{equation}
\Delta^2=C_c \left( Re(A_2/{\cal B})-Re(A_1/{\cal B})
\frac{Im(A_2/{\cal B})}{Im(A_1/{\cal B})} \right) \quad 
\end{equation}
is a typical amplitude.  We have thus demonstrated that
Eq. (\ref{eq:nonlin}) characterizes a Hopf-bifurcation if the complex
coefficient ${\cal A}$ vanishes at a critical point $C_c$ for a
critical frequency $\omega_c$.

\subsection{Amplified response to sinusoidal stimuli}

If a sinusoidal stimulus $f(t) = f_1 e^{i \omega t} + f_{-1}
e^{-i\omega t}$, for which all $f_n$ with $n\neq \pm 1$ vanish,
Eq. (\ref{eq:nonlin}) becomes
\begin{equation}
f_1 \simeq {\cal A} x_1 + {\cal B} |x_1|^2 x_{1} \quad . \label{eq:fx}
\end{equation}
We consider a system that is tuned exactly to the bifurcation,
$C=C_c$. In this situation spontaneous oscillations do not occur and
${\cal A}= (\omega-\omega_c) A_1$.
If the imposed frequency is equal to the critical frequency
$\omega=\omega_c$, the coefficient ${\cal A}$ vanishes and we can
solve Eq. (\ref{eq:fx}) for $|x_1|$ to find the nonlinear response
\begin{equation}
|x_1|\simeq |{\cal B}|^{-1/3} |f_1|^{1/3} \quad , \label{eq:xfnl}
\end{equation}
as a function of the force amplitude $|f_1|$. This behavior represents
an amplified response with a gain
\begin{equation}
r=\frac{|x_1|}{|f_1|} \sim |f_1|^{-2/3} \quad 
\end{equation}
that becomes arbitrarily large for small forces. If the frequency
$\omega$ is different from $\omega_c$, this nonlinear response still
holds as long as the linear term in Eq. (\ref{eq:fx}) is small
compared to the cubic term and can be neglected. This is the case if
$|x_1|^2 \gg |{\cal A}/{\cal B}| = |\omega-\omega_c||A_1/{\cal
B}|$. Therefore, the nonlinear regime characterized by
Eq. (\ref{eq:xfnl}) holds for sufficiently large force amplitudes,
$|f_1|\gg |(\omega-\omega_c) A_1|^{3/2}/|{\cal B}|^{1/2}$, or if the
frequency is sufficiently close to the critical frequency,
$|\omega-\omega_c|
\ll |f_1|^{2/3}|{\cal B}|^{1/3}/|A_1|$. 

If the frequency mismatch $|\omega-\omega_c|$ becomes large, or if
forces $|f_1|$ are small, a new regime occurs for which the linear
term in (\ref{eq:fx}) dominates. In this regime, the response is
linear, 
\begin{equation}
|x_1| \simeq  \frac{|f_1|}{|(\omega-\omega_c)A_1|} \quad ,
\end{equation}
and the gain is constant. This is a passive response if the stimulus
frequency is too far from the critical frequency.

\subsection{Additional remarks}

The above derivation is based on an expansion
(\ref{eq:expnl}) in the displacements $x_n$. This excludes some
nonlinearities in the force which can lead to additional nonlinear
terms in Eq. (\ref{eq:fx}). The most general form of Eq. (\ref{eq:fx})
is
\begin{eqnarray}
f_1 \simeq {\cal A} x_1 &+& {\cal B} |x_1|^2 x_{1} + \ {\cal C} x_1
|f_1|^2 + {\cal D} x_{-1} f_1^2 \nonumber \\ 
&+& {\cal E} |x_1|^2 f_1 + {\cal F}
x_1^2 f_{-1} + {\cal G} |f_1|^2 f_1 \quad . \label{eq:fx2}
\end{eqnarray} 
However, for small forces $f_1$ and small amplitudes $x_1$, the
results derived above are not affected. The regime of nonlinear
reponse $|f_1|\sim |x_1|^{3}$, as well as the linear response regime
$|f_1|\sim |x_1|$ still exist. If $|f_1|\sim|x_1|$, the nonlinear
terms in $f_1$ renormalize the third order term, which in this regime
is negligable. If $|f_1|\sim |x_1|^3$, the nonlinear terms in
$f_1$ are of even higher order and can be neglected.

\section{Oscillations generated by molecular motors}

\subsection{Two state model}

The two state model describes force-generation as a result of
transitions between two states, a bound state and a detached state of
a motor and its track filament. The interaction between a motor at
position $z$ along the filament in states $1$ and $2$ is characterized
by two periodic potentials $W_1(z)=W_1(z+l)$ and $W_2(z) =W_2(z+l)$
where $l$ is the period. We introduce the relative position $\xi=z \;
{\rm mod} \; l$ with respect to the potential period.  Detachment and
attachment rates are denoted $\omega_1(\xi)$ and $\omega_2(\xi)$,
respectively. Oscillations can occur in this model if a large number
$N$ of motors move collectively against an external elastic element of
modulus $K$.

We introduce the probability $P_1(\xi)$ and $P_2(\xi)$ of finding a motor
bound at position $\xi$ in state $1$ or $2$, which satisfy the
normalization condition 
\begin{equation}
\int_0^l d\xi (P_1+P_2) = 1
\end{equation}
For a large number of motors collectively moving with the same
velocity $v$ the dynamic equations read
\begin{eqnarray}
\partial_t P_1 + v \partial_\xi P_1 & = & -\omega_1 P_1 + \omega_2 P_2 \\
\partial_t P_2 + v \partial_\xi P_2 & = & \omega_1 P_1 - \omega_2 P_2 
\end{eqnarray}
The velocity $v$ is determined by the force-balance condition
\begin{equation}
f = \lambda v + K z + N \int_0^l d\xi (P_1 \partial_\xi W_1 + P_2
\partial_\xi W_2)
\end{equation}
where $\lambda$ is a friction coefficient describing the total
friction and $z$ is the displacement of the motors, $\partial_t z=v$.
For an incommensurate arrangement of motors with respect to the track
filament and a large number $N$ of motors, $P_1(\xi)+P_2(\xi)=1/l$ and
the equations of motions simplify:
\begin{equation}
\partial_t P + v \partial_\xi P = -(\omega_1+\omega_2) P + \omega_2/l
\label{eq:P1} \quad ,
\end{equation}
where we denote for simplicity $P(\xi)=P_1(\xi)$.

We discuss a simple choice for the potentials and transition rates for
which the Hopf bifurcation is easy to determine analytically.
We consider the potential 
\begin{equation}
W_1(\xi)= U \cos(2\pi \xi/l)\label{eq:W1}
\end{equation}
with amplitude $U$, and the potential $W_2$ to be constant. The
transition rates are chosen to be periodic functions
\begin{eqnarray}
\omega_1(\xi)&=& \beta - \beta \cos (2 \pi \xi/l) \\
\omega_2(\xi)&=& \alpha - \beta + \beta \cos (2\pi \xi/l) \label{eq:w2}
\end{eqnarray}
parameterized by two coefficients $\alpha$ and $\beta$. With this
choice, 
\begin{equation}
\omega_1(\xi)+\omega_2(\xi) = \alpha
\end{equation}
is constant and the fact that $\omega_1$ and $\omega_2$ are positive
restricts $\beta$ to the interval $0 \leq \beta \leq \alpha/2$.

\subsection{Linear response function}

In order to determine the linear coefficient ${\cal A}$ which
determines the stability of the system, we look for small amplitude
oscillations close to the resting state with $v=0$.  We write
\begin{eqnarray}
P&\simeq &p_0 + p_1 e^{i\omega t} \\
f&\simeq & f_1 e^{i\omega t}\\
z&\simeq & z_1 e^{i\omega t}
\end{eqnarray}
where $p_0=\omega_2/\alpha l$.  To linear order in $z_1$, we find from
Eq. (\ref{eq:P1})
\begin{equation}
p_1=-\frac{i \omega z_1}{i\omega+\alpha}\partial_x p_0
\end{equation}
The corresponding force is given by
\begin{equation}
f_1 \simeq {\cal A} z_1
\end{equation}
with
\begin{equation}
{\cal A} = i \omega\lambda + K + \chi \quad ,
\end{equation}
where the active response $\chi$ of the motors is given by
\begin{equation}
\chi= - N \int_0^l d\xi \frac{i\omega} 
{i\omega+\alpha}\partial_\xi p_0 \partial_\xi W_1
\end{equation}
For the choice of Eq. (\ref{eq:W1}) and (\ref{eq:w2}) the
integral can be calculated and we obtain
\begin{equation}
{\cal A}(C,\omega) = i\omega \lambda +K - N k_0 C \frac{i \omega/\alpha +
(\omega/\alpha)^2}
{1+(\omega/\alpha)^2} \quad . \label{eq:Acal}
\end{equation}
Here, we have introduced the dimensionless control parameter $C \equiv
2\pi^2\beta/\alpha$ with $0<C<\pi^2$ and the cross-bridge elasticity
$k_0\equiv U/l^2$ of the motors.

\subsection{Hopf bifurcation}

A Hopf bifurcation occurs if there is a pair of values $(C,\omega)$
for which ${\cal A}$ as given by Eq. (\ref{eq:Acal}) vanishes. Such a
point indeed exists. For the critical value 
\begin{equation}
C_c = \frac{\lambda\alpha+K}{N k_0}
\end{equation}
the bifurcation occurs for the critical frequency
\begin{equation}
\omega_c = \left(\frac{K\alpha}{\lambda}\right)^{1/2}
\end{equation}
The critical frequency is bounded by the fact that $C_c<\pi^2$.
The maximal frequency occurs for the maximal possible value
of $K$ 
\begin{equation}
K_{\rm max} = N k_0 \pi^2-\lambda\alpha
\end{equation}
for which $C_c=\pi^2$. This frequency is given by
\begin{equation}
\omega_{\rm max}= \alpha \left( N \pi^2\frac{k_0}{\lambda\alpha}
- 1\right)^{1/2}
\end{equation}
Note, that the maximal frequency can be significantly higher than the
typical rate $\alpha$ of the chemical cycle.

\end{document}